# Performance Enhancement of Soil-Structure Systems Using a Controlled Rocking

Aria Fathi[1], S. Mohsen Haeri, Ph.D., P.E., D.I.C. [2], Mehrdad Palizi[3], Mehran Mazari, Ph.D.[4], Cesar Tirado, Ph.D. [1], Cheng Zhu[1]

[1]Center for Transportation Infrastructure Systems (CTIS), The University of Texas at El Paso, 500 W. University Ave., El Paso, TX 79968; e-mail: afathi@miners.utep.edu, ctirado@utep.edu, czhu@miners.utep.edu

[2]Department of Civil Engineering, Sharif University of Technology, Tehran, Iran; e-mail: smhaeri@sharif.edu

[3]Department of Civil Engineering, The University of Alberta, 116 St. and 85 Ave., Edmonton, AB, Canada; e-mail: palizi@ualberta.ca

[4]Department of Civil Engineering, California State University Los Angeles, 5151 State University Drive, Los Angeles, CA 90032; e-mail: mmazari2@calstatela.edu

**ABSTRACT**

Application of performance-based design (PBD) in earthquake geotechnical design codes has been gaining attention for the past decade. Strong vibrations (e.g., earthquake loading) may generate an uplift or a partial separation of shallow foundations from the underneath soil. To minimize the damage and keep the superstructure safe from the vibrations, a controlled rocking can be exploited. In this study, a finite element method is used to investigate the effect of nonlinear behavior of foundation geomaterial on the rocking behavior of soil-structure systems. To get a better insight about the rocking behavior of the soil-structure system, nonlinear elastic-perfectly plastic behavior is considered for the simulated geomaterial of the foundation. Next, a parametric study, using a wide range of geomaterials with different stiffness values, is conducted in accordance with high-rise structures with different configurations. The results indicate that a significant proportion of the input energy is dissipated by deploying rotational behavior of shallow foundations especially for the structures on softer geomaterials.

## 1. Introduction

Strong seismic motion causes large inertial and eccentric forces to the slender and high-rise structures. The supporting shallow foundations beneath the structures may experience a partial separation from the underneath soil due to the overturning moments. The uplift of one side of the foundation is simultaneously accompanied by an increase in the shear stress on the opposite side of the foundation; therefore, a failure mechanism may happen as the combined shear and normal stresses exceed shear strength of the geomaterial, which might be associated with permanent displacement of the foundation. Many attempts have been focused on predicting the behavior of the geomaterial during foundation rocking as an established tool for performance enhancement of buildings during severe vibrations (Housner 1963; Koh et al. 1986; Gottardi and Butterfield 1995; Zhang and Makris 2001; Gajan et al. 2005; Adamidis et al. 2014; Fathi 2014; Abadi et al., 2015; Taghavi et al. 2015; Haeri and Fathi 2015; Shahir et al. 2016; Pak et al. 2016, Sharma and Deng 2017, Barari et al. 2017; Sharma et al. 2018).

One of the main transitions from traditional geotechnical earthquake design to performance-based design (PBD) is to emphasize on soil-structure interaction (SSI) (Keshavarzi and Bakhshi 2012; Rahimi et al. 2018; Lemus et al. 2018). The new practices are thus involved in design concepts to consistently achieve a desired level of satisfaction in performance enhancement of structures during a severe earthquake. Different approaches have been proposed different performance criteria, e.g. Ayoubi and Pak (2017) considered the settlement as a performance criterion and were able to develop a practical formula in order to calculate the settlement of an structure. PBD is normally justified on the basis of cyclic components of motion as accepted in the traditional design codes (e.g., FEMA 2000). The study conducted by Anastasopoulos et al. (2010) showed that allowing a footing to yield, instead of only the superstructure, can improve the overall integrity of the structure during an earthquake. Plastic "hinging" of foundation is an instance of such behavior. Figure 1 shows the enhancement in overall performance of the structure (the structure is not collapsed) due to foundation hinging (Hume, 2017) which happened as a result of basin effects in Kathmandu Valley, Nepal during the $M_w$ 7.8 Gorkha 2015 earthquake (Ayoubi et al, 2018).

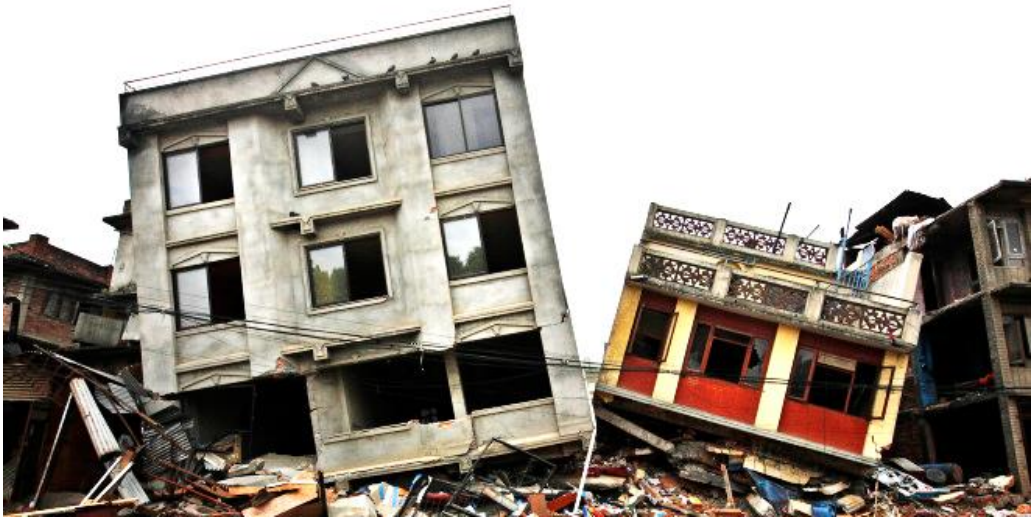

**Figure 1.** Plastic "hinging" of foundation due to the rocking mechanism in Kathmandu Valley, Nepal as a result of the $M_w$ 7.8 2015 Gorkha earthquake (Hume, 2017).

Another study by Liu et al. (2013) concluded that the input energy to the soil-structure system can be dissipated between superstructure and foundation by a balanced design strategy. Figure 2 demonstrates the schematic of the two above-mentioned concepts, i.e., foundation failure and structural failure.

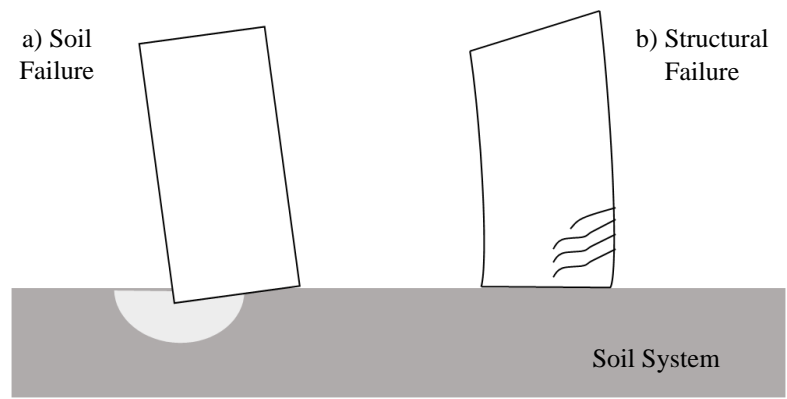

**Figure 2.** Schematic of design concepts; a) Plastic "hinging" at the soil and b) Structural failure.

Gajan and Kutter (2008) performed several centrifuge tests to evaluate the rocking behavior of shallow foundations attached to shear wall structures and supported by sandy and clayey materials during slow lateral cyclic loading. They concluded that partial separation of foundation and the corresponding soil yielding, caused by foundation rocking, can be considered as energy dissipation mechanisms for the soil-superstructure system. The beneficial effects of rocking and cyclic load-displacement of shallow foundations have been studied in several experiments at low confinement stresses (1g acceleration by shake table) as documented in Meek (1975), Bartlett (1976), Xia and Hanson (1992), Martin and Lam (2000), Pecker and Pender (2000), Faccioli et al. (2001), Jafarzadeh et al. (2013); Haeri et al. (2015); and at high confinement stresses (ng acceleration by centrifuge) as reported in Garnier et al. (2007), Gajan and Kutter (2008), Deng et al. (2012).

Aside from the benefits of rocking in seismic design codes, there is still a gap for deploying the above-mentioned advantages to practice. Thus, the ability of realistically simulating the rocking response of the foundations is a necessity in this respect. A progressed numerical algorithm including a well-matched constitutive model is clearly required to simulate the behavior of soil system. Even though complex constitutive models can be found in several studies (e.g. Kutter 2006; Anastasopoulos et al. 2010; Abate et al 2010; Anastasopoulos 2011; Panagiotidou et al. 2012; Coe et al. 2016; Keshavarzi and Kim 2016; Ashtiani et al. 2017; Sebaaly et al. 2017; Mahvelati and Coe 2017; Tirado et al. 2017; Barari et al. 2017; Rashidi and Haeri 2017; Rashidi et al 2017; and Fathi et al. 2018), it expected that introducing nonlinearity in the simulated geomaterials will produce more practical and reliable results to predict the rocking behavior of the foundations (Haeri and Fathi 2015). Nonlinear behavior of geomaterials triggers a probable increase in the natural period of the soil-structure system and a decrease in shear base demand in structures during seismic loading (Mylonakis and Gazetas 2000).

This study attempts to provide strategies to integrate soil-structure interaction and rocking behavior in geotechnical design and practice. A two-dimensional finite element (FE) model, that simulates the interaction between the soil and the foundation, is developed to investigate the rocking behavior of soil-structure systems induced by slow cyclic loading. The FE model adopted nonlinear properties of the geomaterials for conducting a parametric study. The effect of different parameters (e.g., building height, rotation, and soil stiffness) on energy dissipation and rocking responses of the foundation are evaluated. The following sections discuss the details of FE models followed by the results and discussion.

## 2. Finite Element Modelling of Soil-Structure System

The rocking behavior of the soil-structure system was simulated by implementing **ABAQUS®**—a multi-purpose FEM program—that uses explicit and implicit time integration techniques. A 2D dynamic FE model was assembled to simulate a superstructure imparting energy to the geomaterial at a given amplitude and frequency. The simulated soil-structure system is shown in Figure 3. The assembled strip foundation-superstructure system was simulated using plain strain elements. Element-type CPE8R—8 nodes biquadratic, reduced integration—was used throughout the model. Strip foundations were supported by a sandy soil stratum. Even though the soil system was simulated large enough using CPE8R elements with the dimensions shown in Figure 3, non-reflective boundaries were taken into consideration to prevent the ongoing reflected waves. The mesh was finely discretized with 10 elements under the length of the foundation. Smaller elements were used for the geomaterials beneath the superstructure, and larger elements were used beyond that.

Unlike Hertzian models and iterative methods in boundary element models, surface to surface contact models implemented in the FEM model, offer the advantage of capturing the effect of nonlinearity of geomaterials on foundation contact area. In FEM, the contact area is simulated by using algorithms that search for penetration of a slave node into a master segment. General surface-to-surface contacts further allow decoupling of both surfaces by releasing the slave node and setting contact forces to zero. These conditions are suitable for the modeling of a rocking behavior of foundation that can be separated from the underneath geomaterials. The interaction between foundation and soil system was then considered using ABAQUS's general contact model. The coefficient of friction ($\mu$) was assumed 1.25 to prevent undesired sliding.

To simulate stress states in the geomaterial during rocking of shallow foundations, Mohr-Coulomb model was employed in this study. There is a consensus among researchers that shear modulus changes with the variation in shear strain and confining pressure (e.g., e.g., Seed and Idriss 1970; Shibata and Soelarno 1975). In other words, shear modulus increases with an increase in confining stress and decreases with an increase in strain level. However, in most of dynamic SSI analyses, the plasticity of geomaterial is not considered. In this study, through an iterative process, the geomaterial modulus was adjusted with the strain and stress levels during the FE analysis. Hence, the soil system under the application of static and slow cyclic loadings behaves nonlinearly before the Mohr-Coulomb failure envelope, and thereafter behaves plastic by reaching the shear strength (i.e., Mohr-Coulomb failure envelope).

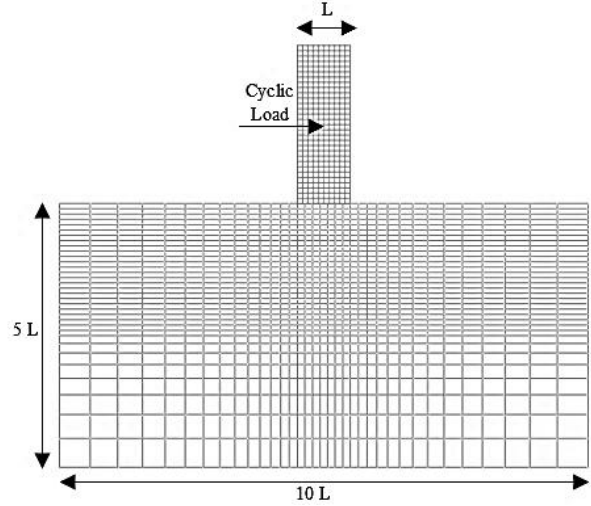

**Figure 3.** FE model of the soil-structure system.

A nonlinear model proposed by (Seed and Idriss 1970; and Seed et al. 1986) was found to be as a proper indicator for representing the dynamic shear moduli of granular soils. The nonlinear model is as follows:

$$G_{max} = 218.82 \, K_{2(max)} (\sigma')^{0.5} \tag{1}$$

where $G_{max}$ is maximum shear modulus, $K_{2(max)}$ is a laboratory shear modulus coefficient measured at low strain level, and $\sigma'$ is mean effective principal stress which is defined as:

$$\sigma' = \frac{\sigma'_1 + \sigma'_2 + \sigma'_3}{3} \tag{2}$$

where $\sigma_i$ are the principal stresses. The ratio of shear modulus to the maximum shear modulus at different strain levels ($F' = \frac{G}{G_{max}}$) was suggested by Seed et al. (1986) as depicted in Figure 4. The magnitude of shear modulus was therefore adjusted at different strain levels during FE simulations.

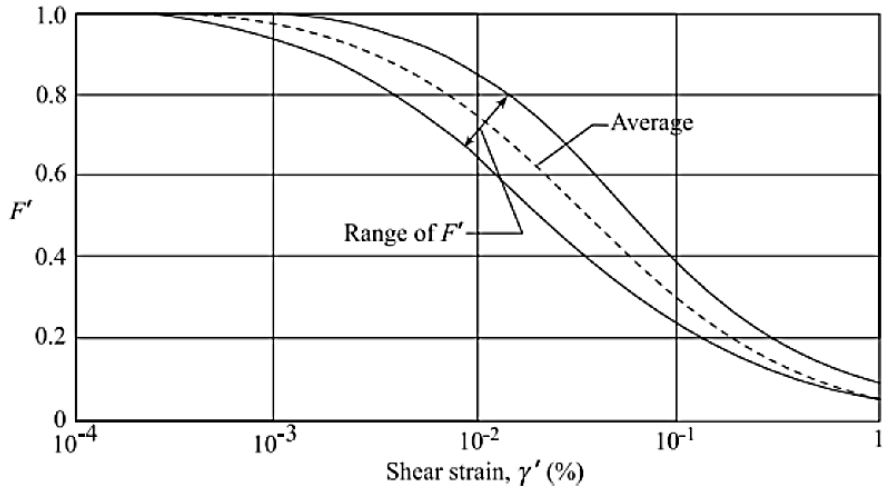

**Figure 4.** Shear modulus variation at different strain levels (After Seed et al. 1986).

Rayleigh damping (Rayleigh and Lindsay 1945) was introduced to the soil-structure system as:

$$[C] = \alpha[M] + \beta[K] \quad (3)$$

where $[M]$ is the mass matrix, $[K]$ is the stiffness matrix, and Rayleigh constants are $\alpha$ and $\beta$, which were computed using the following equation:

$$2\xi\omega_i = \alpha + \beta\omega_i^2 \quad (4)$$

where $\xi$ is damping ratio; and $\omega_i$ is natural frequency of the $i^{th}$ mode. Several studies have been implemented to estimate the damping ratio of buildings in recent years (Londono and Neild 2012; and Bernal et al. 2015). Bernal et al. (2015) derived the first mode damping ratio for over 200 buildings with different heights (being steel or concrete structures). The authors showed that the damping ratio can change from 3 to 7% for the structures less than 100 meters height. For this study, $\xi$ was assumed as 8% and 5% for the soil and the superstructure systems, respectively. $\alpha$ and $\beta$ can be calculated using two values of natural frequencies obtained from two different modes. Hence, one of the most imperative factors for obtaining Rayleigh damping coefficients is the selection of the two most influential modes (or natural frequencies). Park and Hashash (2004) concluded that the fundamental natural frequency and the first odd mode, for which the frequency is greater than the loading frequency, can be considered to satisfy Eq. 3. Following their conclusion, the Rayleigh constants were determined as $\alpha = 0.214$ and $\beta = 0.03$ for the geosystem in this study.

A comprehensive parametric study is conducted for high-rise structures with different heights and weights built on geomaterials with different properties. The features considered for the simulated superstructures are reported in Table 1. For the soil system, a series of nonlinear $K_{2(max)}$ parameters were randomly selected from a range of values considered for sandy materials recommended by Seed et al. (1986). The feasible ranges of the $K_{2(max)}$ parameter are reported in Table 2.

**Table 1.** Specification for Simulated Structures.

| Structure Features | Symbol | Range of Values |
|---|---|---|
| *Weight* | $W$ | 19.6 – 58.8 MN |
| *Height* | $h_s$ | 30 – 60 m |
| Length | $L$ | 20 m |

A set of 400 soil-structure systems were simulated within the range of values in Tables 1 and 2. For each case, a combination of superstructure and substructure (very loose to very dense sandy materials, $30 < K_{2(max)} < 75$) was randomly selected.

**Table 2.** Feasible Range of Geomaterial Properties (Seed et al. 1986).

| Geomaterials Properties | Range of Values |
|---|---|
| $K_{2(max)}$ | 30 – 75 |
| *Cohesion (C)* | 15 kPa |
| *Angle of internal friction ($\varphi$)* | 38 |
| *Dilatation angle* | 3 |
| *Poisson's ratio* | 0.3 |

The assembled foundation-superstructure system was subjected to an application of slow lateral cyclic loading (displacement-controlled) at its center of gravity of superstructure. Figure 6 shows the applied time history of slow cyclic loading for the assembled foundation-structure systems. The displacement-time history contains three clusters of cyclic loading with different amplitudes. Therefore, the applied sinusoidal lateral displacement provides the shallow foundation with three ranges of rotations through the time history of loading. The applied rotations are 0.0015, 0.005, and 0.015 radians.

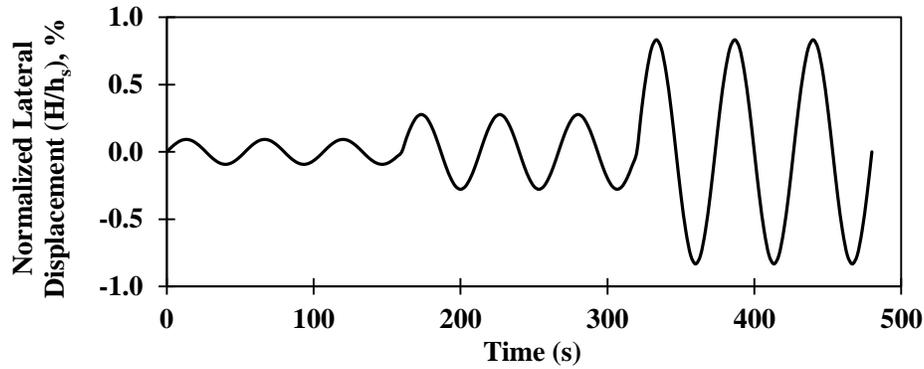

**Figure 6.** Time history of slow cyclic loading.

## 3. Model Validation

The preliminary validation of simulated FE models was performed using centrifuge test results reported by Rosebrook and Kutter (2001). A double-wall structure configuration comprised of two aluminum shear walls with parallel aluminum strip footings sitting on Nevada sand with a relative density of 60 percent was tested at 20g centrifugal acceleration. Table 3 summaries some of the properties of Nevada sand used for the centrifuge tests. Rosebrook and Kutter (2001) used a viscose material on the surface of Nevada sand to protect the curvature of the building footprint during foundation rocking. The angle of repose and cohesion of the simulated sandy materials were back calculated using the failure envelopes acquired from shear box tests (see Figure 7). These parameters were calculated as $\varphi = 33°$ and $C = 10$ kPa.

**Table 3.** Material Properties of Nevada Sand (Rosebrook and Kutter 2001).

| Geomaterials Properties | Values |
|---|---|
| Classification | Uniform, fine sand; SP |
| Specific Gravity | 2.67 |
| Mean Grain Size, $D_{50}$ (mm) | 0.15 |
| Coefficient of Uniformity, $C_u$ | 1.06 |
| Maximum Dry Unit Weight, $\gamma_{d,\,max}$ (kN/m³) | 16.8 |
| Minimum Dry Unit Weight, $\gamma_{d,\,min}$ (kN/m³) | 14.0 |
| Relative Density (%) | 60 |

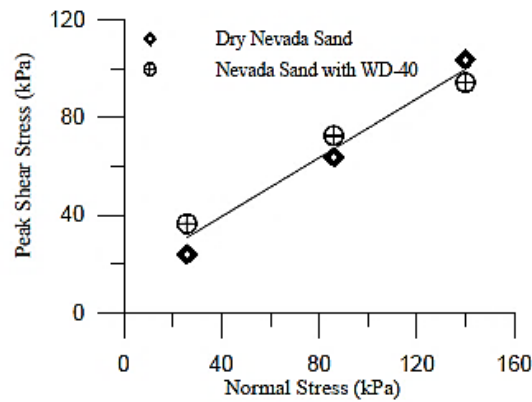

**Figure 7.** Failure envelopes obtained from shear box test (After Rosebrook and Kutter 2001)

Schematic of the assembled centrifuge model is depicted in Figure 8a. The model was subjected to cyclic loading, i.e. a controlled displacement applied by an actuator. The time history of applied lateral displacement is shown in Figure 8b. The scale factors provided by the authors were used to convert the scale model to prototype units. The length, width, and the height of the shear wall in the prototype scale were determined to be 2.68 m, 0.68 m and 10.14 m, respectively. A more detailed explanation of the material properties and the assembled models can be found in Rosebrook and Kutter (2001). The simulated FE model of the shear-wall structure is illustrated in Figure 9. Due to axisymmetric properties of the system, only a half-space model of the double-wall structure was simulated.

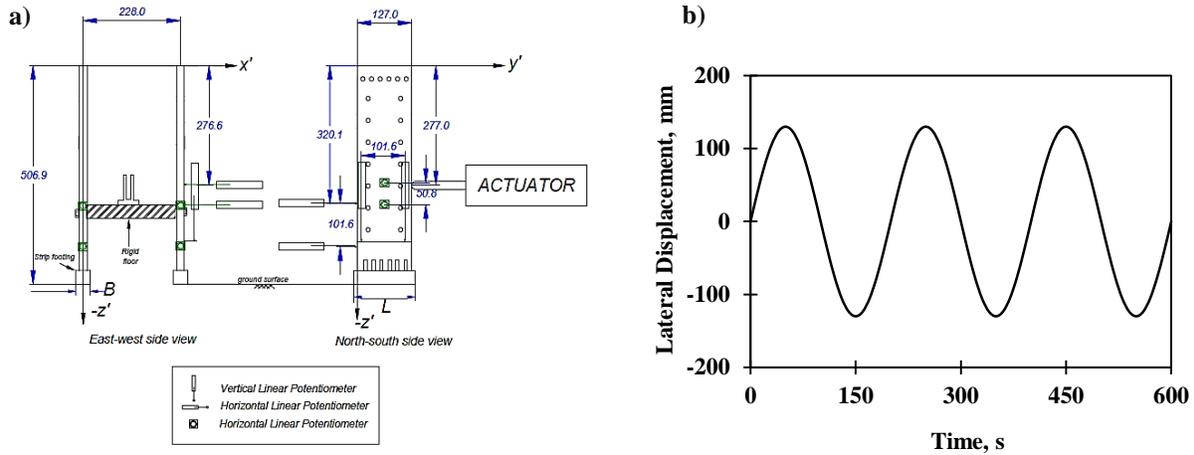

**Figure 8.** a) Schematic view of assembled centrifuge model under lateral cyclic loading and the assembled shear wall structure; all units are in model scale millimeters (Rosebrook and Kutter 2001). and b) Time history of lateral displacement.

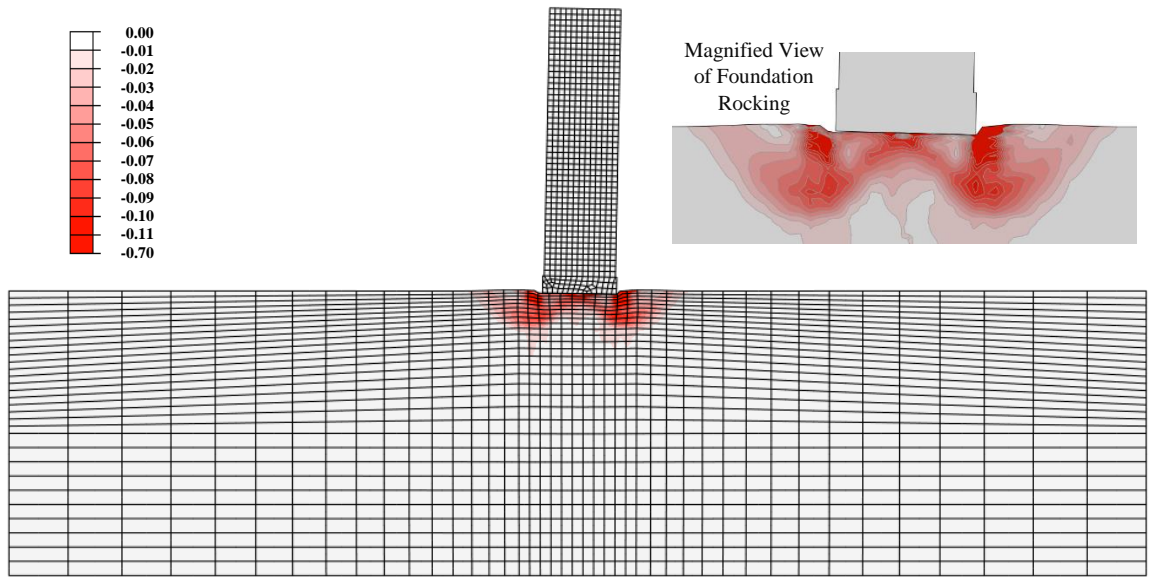

**Figure 9.** Simulated FE model of Rosebrook and Kutter (2001) experiment showing plastic strain contours in principal plane on a deformed mesh.

The behavior of geomaterial was assumed nonlinear elastic-perfectly plastic; while, the wall structure behaves linearly elastic deliberating aluminum elastic modulus in the analysis. The modulus of geomaterials was set to be variable in the model. During application of lateral displacement, gap formation can be governed on one side of the foundation; while, the other side causes an increase in shear stress on supporting material. The increase may result in yielding of the material at the opposite side of the foundation. The material plasticity is demonstrated using plastic strain contours on deformed elements in Figure 9.

Figure 10 shows the rotational behavior of the soil-structure system (being rotation-moment and rotation-displacement) at the center point of the foundation. Rotation-moment relationship encloses a large area for both centrifuge test and FE analysis, indicating a significant amount of dissipated energy (Figure 10a). The dissipation of energy is associated with yielding of the soil beneath the structure. Permanent settlement of the foundation, due to soil plasticity, can be observed for both centrifuge and FE models in Figure 10b. The gathered results from the FE model were appreciably promising in terms of rotation-moment and rotation-displacement, showing that the implemented FE model is potentially successful in prediction of rocking behavior.

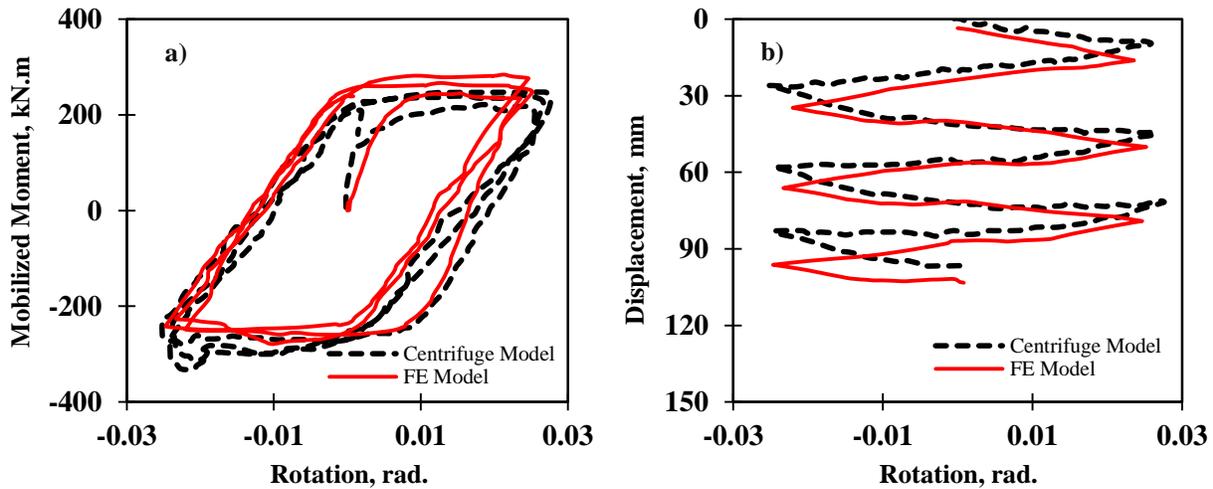

**Figure 10.** Model validation using centrifuge test results from Rosebrook and Kutter (2001) experiment; a) Rotation-mobilized moment relationship, and b) Rotation-displacement relationship.

## 4. Results and Discussion

### 4.1 Permanent Settlement

The aim of this part of the study is to evaluate the effect of soil stiffness and vertical load (structure weight) on rocking behavior of shallow foundations. The results of different FE models are described in the following sections.

***Vertical load.*** Figure 11 compares initial and permanent displacements due to vertical and lateral loading, respectively. The permanent displacement of the foundation was recorded after each cluster of loading. Three structures with 30, 60, and 90 m heights were selected. These structures were supported by a soil with $K_{2(max)} = 52$ (medium sand). It is shown that the permanent displacement increases significantly with an increase in the vertical load (i.e. structure height). In other words, the taller the structure, the higher permanent displacement during rocking with the same amount of rotation angle would become.

***Stiffness and rotation.*** Considering the nonlinear elastic-perfectly plastic behavior of the underneath geomaterial, the initial and permanent displacements of the shallow foundations after applied rotations were compared. For the purpose of assessing impact of soil stiffness on the foundation responses, the analyzed soils are classified in three ranges of $K_{2(max)}$ values, corresponding to loos (30 <$K_{2(max)}$< 45), medium (45 <$K_{2(max)}$< 60), and dense sandy materials (60 <$K_{2(max)}$< 75).

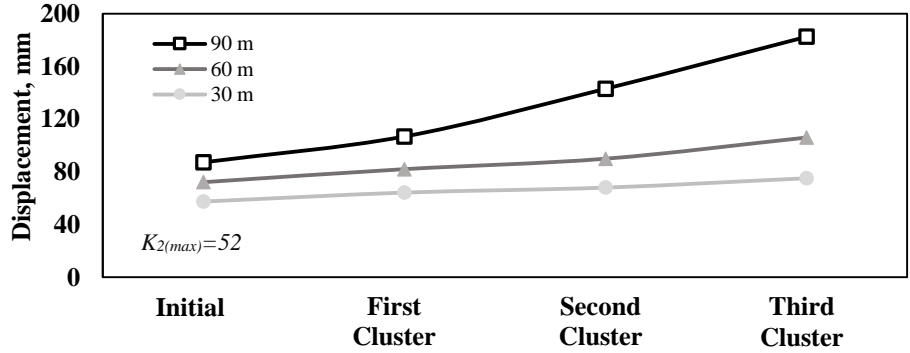

**Figure 11.** Geomaterials initial and permanent displacement after different clusters of lateral loading.

As shown in Figure 12, the initial and permanent displacement values seemed to be well correlated for all the rotations as judged by the number of data points within 20% uncertainty limits. However, an increase in the rotation of the foundation leads to more variations in the results as more scatter is observed for the third cluster of loading (0.015 radians) in comparison to the first (0.0015 radians) and second (0.005 radians) clusters of loading. This is because the soil behaves more plastic under higher loads of the third cluster, while under the first cluster of loading, soil behaves more elastically. The average permanent displacements for first, second, and third clusters are 1.11, 1.25, and 1.43 times that of the average initial displacements, respectively. Figure 12 also illustrates that the displacement decreases as $K_{2(max)}$ increases (material becomes stiffer).

## 4.2 Uplift, Contact Area, and Permanent Displacement

The fundamental theory of contact has been addressed by Hertz (1882) with focusing on the contact area of two elastic objects with curved surfaces. The rocking behavior of foundations can be tightly linked to a contact problem as the applied rotation to the superstructure creates a continuous change in the contact area of the foundation with the underneath geomaterials. The contact area of a rigid foundation and the underneath geomaterials is significantly influenced by the soil stiffness and the type of loading exerted on the superstructure (Gajan and Kutter 2008).

Uplifting of one side of the foundation results in yielding of the geomaterial under the opposite side of the foundation. The permanent displacement of the foundation can be related to the foundation separation and the corresponding foundation-soil contact. This study focuses on assessing the relationship of the above-mentioned responses. The foundation uplift is correlated with permanent displacement as shown in Figure 13. The permanent settlement increases with a decrease in the measured uplift. Also, stiffer soils undergo less permanent displacement as shown in Figure 12 which means higher amount of uplift will be observed for the stiffer material as compared to that for the softer one. The magnitude of uplift at low rotation (0.0015 rad.) is negligible which means that the foundation-soil contact does not change significantly during the first cluster of loading.

High amplitudes of loading (second and third clusters of cycles) exerted on the superstructure generate a partial separation of the foundation from the supporting geomaterials, and thus, the foundation loses a portion of its initial contact with the soil system. The contact area ratio ($\eta$) is defined as the ratio of the contact area ($A_c$) over the actual area of the foundation ($A$) as follows:

$$\eta = \frac{A_c}{A} \quad (5)$$

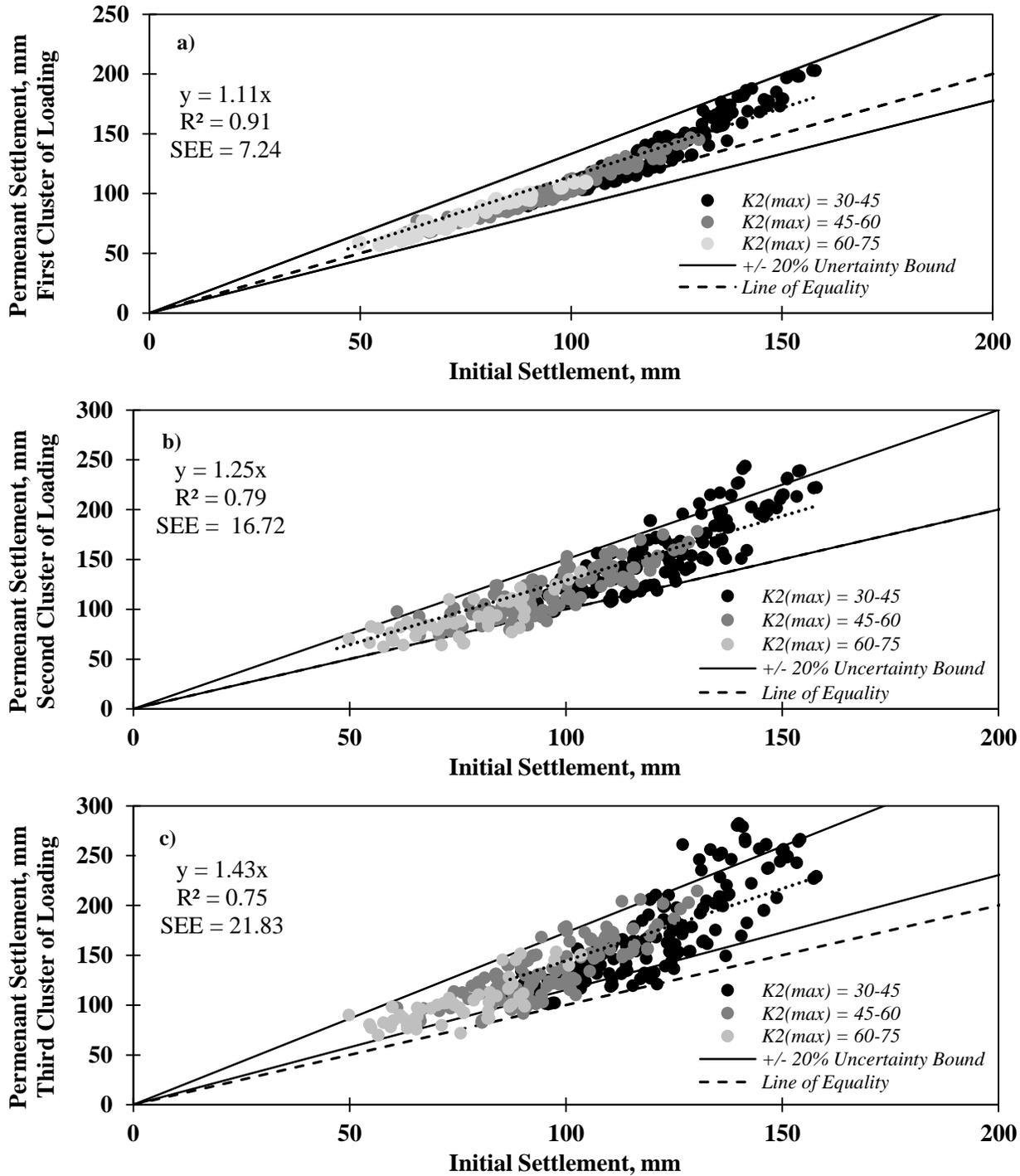

**Figure 12.** Relationship of initial and permanent surface displacements after application of different rotations: a) 0.0015 rad., b) 0.005 rad., and c) 0.015 rad.

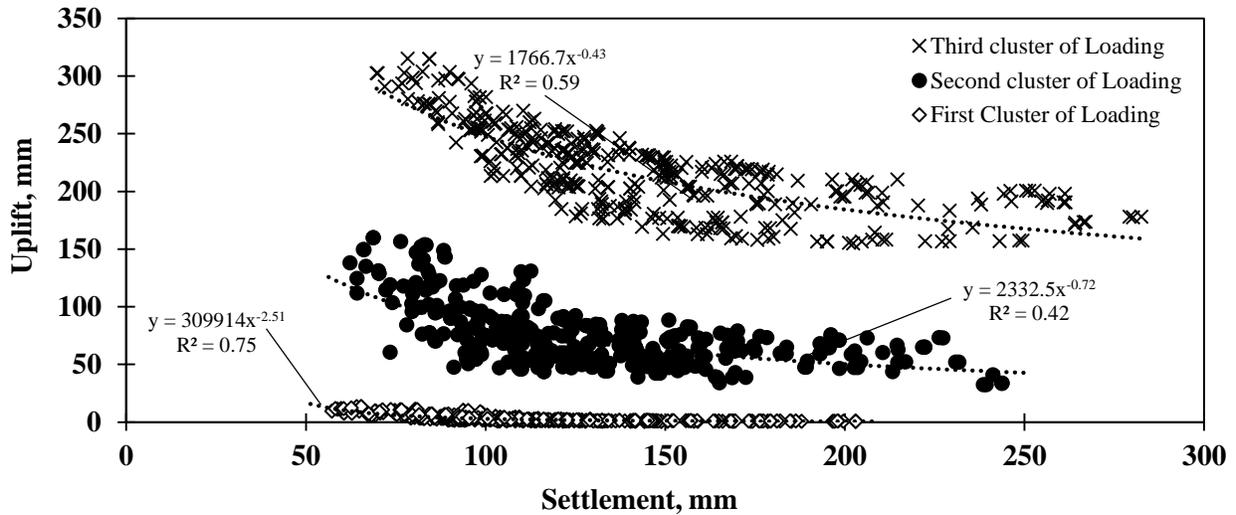

**Figure 13.** Relationship of foundation uplift with the permanent displacements after three clusters of loading.

Figure 14 shows the combined behavior of uplift and contact area, as well as gap formation for different soil stiffness. It is clear that soil stiffness affects both uplift and contact area. Foundation separation on one side is associated with yielding of the soil and closing of the gap on the other side of the foundation. Loose materials can slide and close the gap easier. The foundation is more inclined to lose its contact with the soil-system when it is located on the dense material as compared to the loose one. Therefore, the ratio $A_c/A$ decreases (i.e., foundation loses more contact) with an increase in the stiffness of the geomaterials. By increasing the foundation rocking angle, the contact area ratio ($A_c/A$) decreases which is accompanied by an enhancement in the uplift.

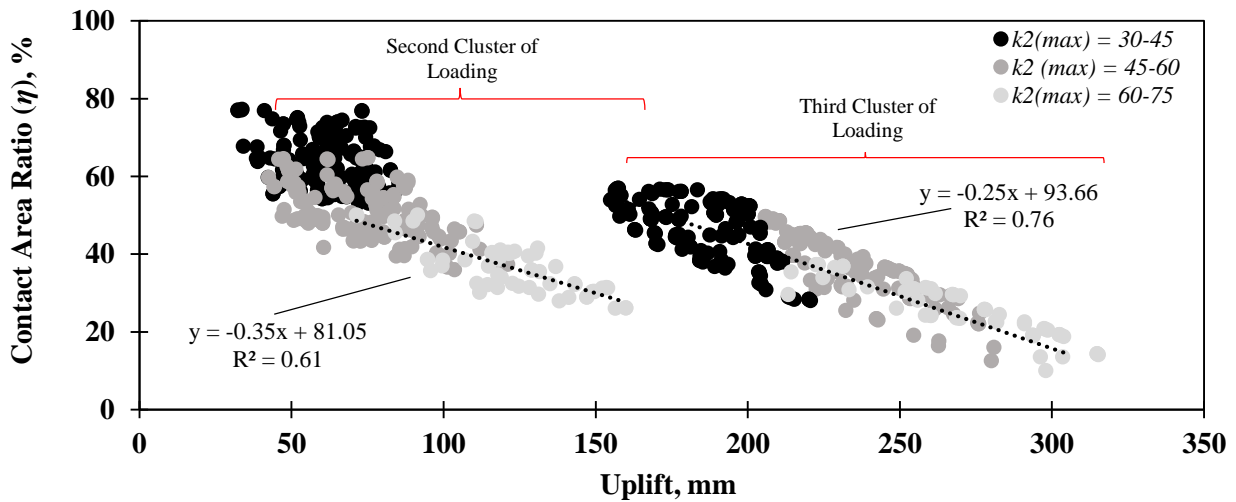

**Figure 14.** Contact area ratio ($\eta$) vs. foundation uplift.

*4.3 Rocking Moment Capacity and Energy Dissipation*

The rocking behavior of shallow foundations as a favorable mechanism to dissipate the exerted input energy on a foundation of superstructure system has been evaluated in different studies in the literature. This paper essentially attempts to evaluate the impact of soil stiffness and vertical loading on energy dissipation and mobilized moment during foundation rocking. Figure 15 depicts a schematic of the soil-foundation-structure system during the lateral slow cyclic load. The free body diagram of forces at the center of the soil-structure interface is also shown in Figure 15. The mobilized moment ($M_{Mobilized}$) about the center of the foundation base can be determined via Equation (6):

$$M_{Mobilized} = F_{react} \cdot h_{cg} \cdot \cos(\theta) - W_s \cdot h_{cg} \cdot \sin(\theta) \tag{6}$$

where $F_{react}$ is the reaction force obtained from the application of lateral cyclic displacement during rocking of the foundation; $h_{cg}$ is the height of the center of gravity of the assembled foundation-superstructure system; $\theta$ is the rotation of the structure due to the lateral cyclic displacement; and $W_s$ is the weight of the structure. Rotation-settlement and corresponding rotation-moment relationships for four FE cases with different soil stiffness under application of different vertical loads (i.e., structures with different weights or heights) are shown in Figure 16. It is observed that foundation rocking causes an accumulation of permanent settlement at the center of the foundation. The permanent settlement gradually increases with an increase in cyclic rotation of the foundation. Two structures with different sizes were supported by a moderately stiff sandy material with $K_{2(max)}$ of 52 as shown in Figures 16a and b. Higher vertical load (weight of structure) results in rapid yielding. Therefore, the foundation accumulates more displacement for the heavier, taller structure (see Figures 16a and 16b). As judged by the enclosed area of the hysteresis loops, the taller buildings located on softer soils are inclined to dissipate more input-energy in comparison to the smaller buildings sitting on the stiffer material. Dissipation of energy is associated with yielding of the soil and accumulation of permanent displacement. Furthermore, energy dissipation increases at higher amplitudes of rotation (hysteresis loops become larger). The ultimate moment capacity, which is accompanied by yielding of the soil beneath the foundation, is mobilized at lower values of rotations (0.5% rad.).

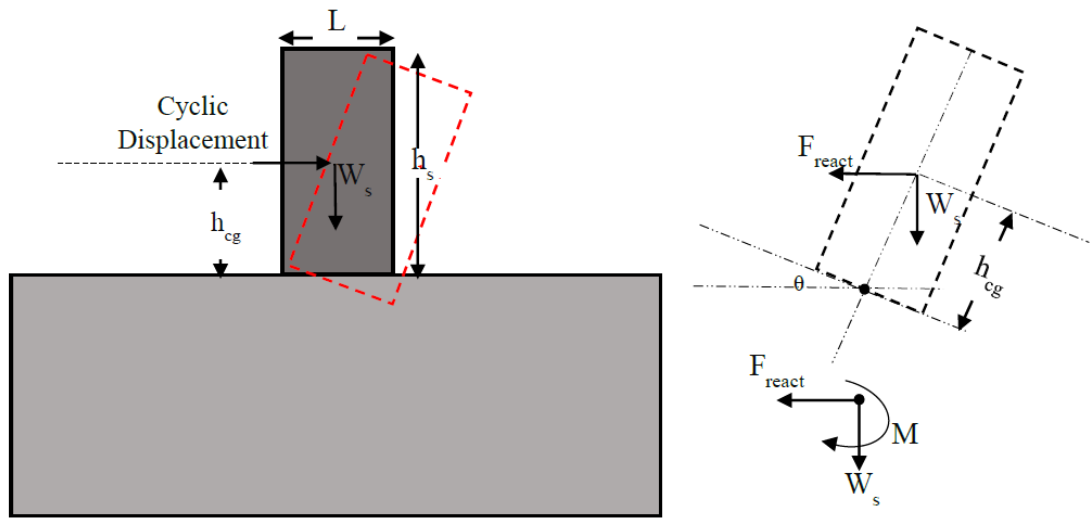

**Figure 15.** Schematic of the soil-structure system and corresponding free body diagram of forces.

Similarly, Figures 16c and 16d demonstrate the rotational behavior of a heavy, high-rise structure supported by dense ($K_{2(max)} = 68$) and loose sandy material ($K_{2(max)} = 40$), respectively. The results show that the foundation displacement per cycle increases significantly for the loose material, and so, more energy is dissipated (see Figure 16d).

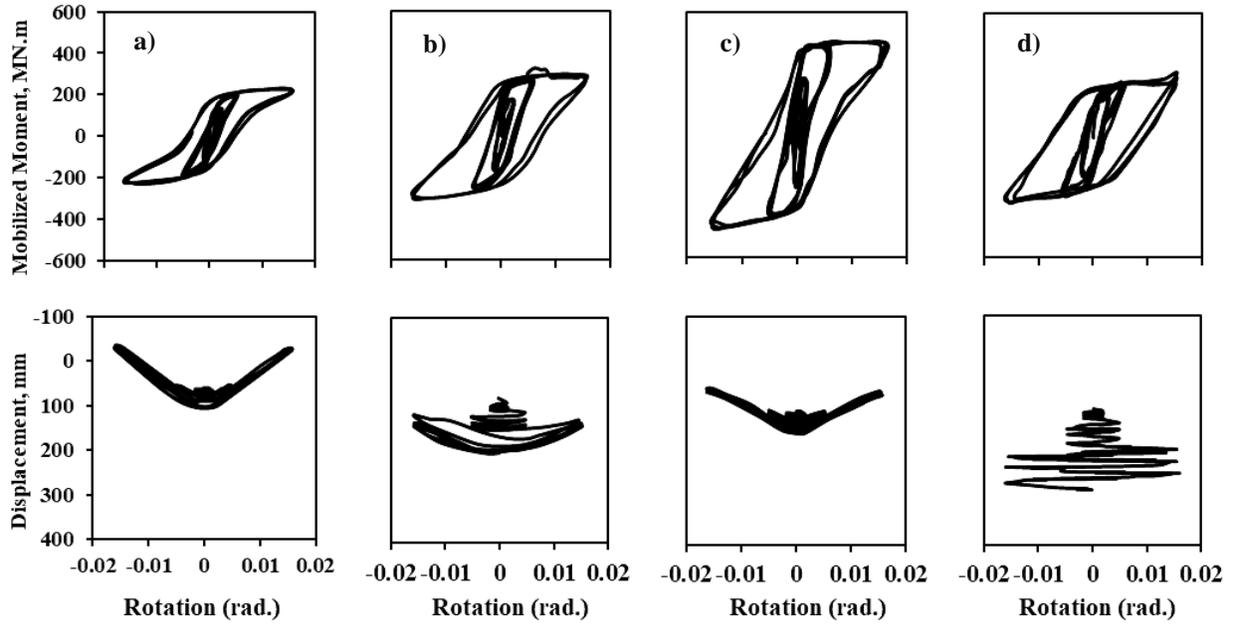

**Figure 16.** Rotation-moment hysteresis loops associated with the corresponding rotation-displacement relationship for soils with different stiffness and structures with different weights: a) $k2_{(max)}=52$, $h_s=32$ m, $W_s=22.38$ MN; b) $k2_{(max)}=52$, $h_s=58.71$ m, $W_s=39.18$ MN; c) $k2_{(max)}=68$, $h_s=87.87$ m, $W_s=57.78$MN; d) $k2_{(max)}=40$, $h_s=87.87$ m, $W_s=57.78$MN.

However, using a stiffer sandy soil with $K_{2(max)}$ of 68, results in more foundation uplift and less dissipation of energy as judged by the area enclosed in the hysteresis loops. The geomaterial stiffness affects the ultimate mobilized moment of the soil-structure system significantly.

In a relevant study, Gajan and Kutter (2008) developed an empirical equation based on a number of centrifuge model tests to calculate rocking moment capacity of a soil-structure system, $M_{c\_foot}$, around the center of the base of a shallow foundation as:

$$M_{c-foot} = \frac{V.L}{2}.(1 - \frac{A_c}{A}) \qquad (7)$$

where $M_{c\text{-}foot}$ is mobilized rocking moment about the base center point of the foundation, $V$ is vertical load (structure weight), $L$ is foundation length, $A_c$ is contact area between soil and foundation during rocking, and $A$ is actual area of the foundation. The contact area ratio ($\eta=A_c/A$) can be back-calculated using Eq. 8 as follows:

$$\eta = \frac{A_c}{A} = 1 - \frac{2.M_{c-foot}}{V.L} \qquad (8)$$

To determine the ratio $A_c/A$, the magnitude of $M_{c\text{-}foot}$ was taken from equation 6. This study compares the ratio $A_c/A$ obtained from the empirical equation with the corresponding results obtained from FE models. The contact area ratio, obtained from FE model and Eq. 8, was compared to the second and third clusters of loading as demonstrated in Figures 17a and 17b, respectively.

The contact area ratio determined from the dynamic FE model is in good agreement and are comparable with the result from Eq. 8 as most of the data points remain within the 20% uncertainty bounds. The average ratio $A_c/A$ for the FE model is about 28% higher than the average ratio obtained from the empirical equation at 0.5% rotation. However, the $A_c/A$ obtained from FE analysis is only 6% lower than that determined from Eq. 8 for 1.5% rotation.

The rocking responses of a soil-structure system including mobilized moment, uplift, permanent displacement, contact area, and energy dissipation have significant impacts on each other. As illustrated in rotation-moment hysteresis in Figure 16, the rotational stiffness decreases as the foundation-superstructure system is subjected to a higher amplitude of lateral loading. The relationship of uplift with the mobilized moment at different rotations is shown in Figure 18. The mobilized moment is highly influenced by the amount of uplift and it increases with an increase in foundation uplift. In other words, more uplift is associated with more eccentricity which results in an enhancement in the mobilized moment.

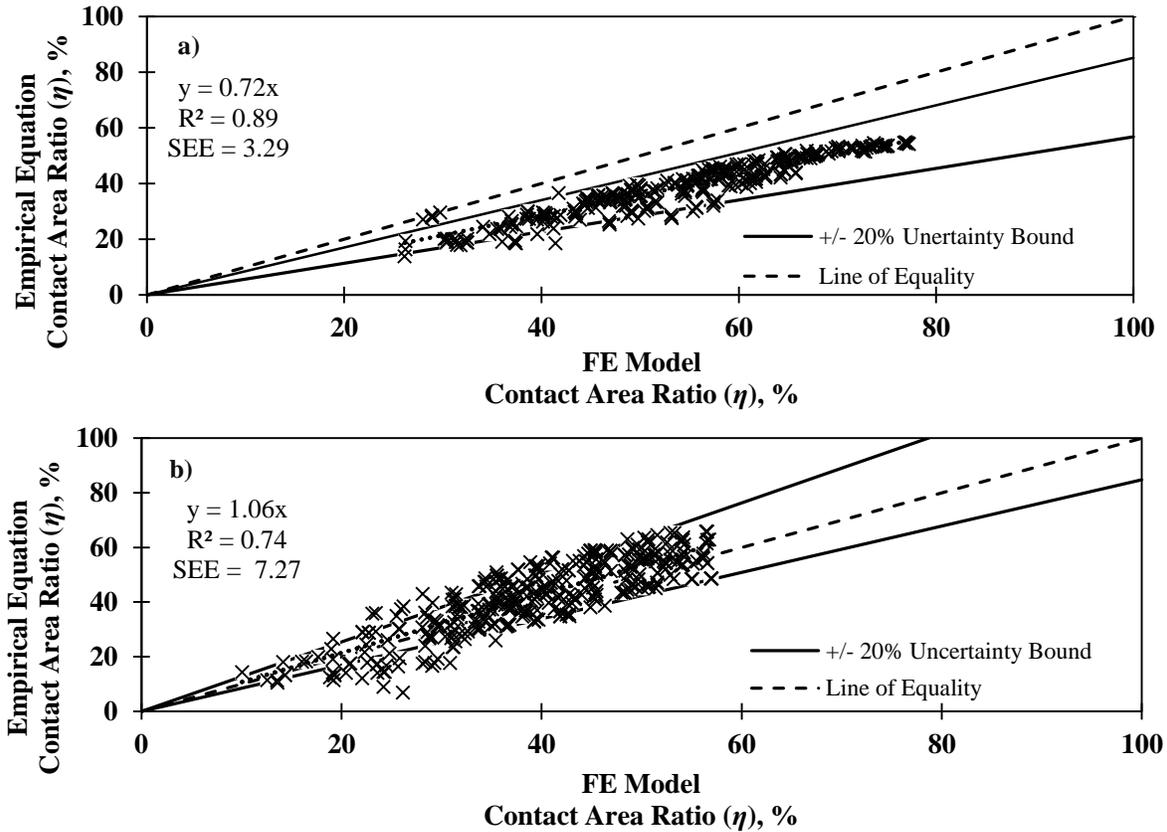

**Figure 17.** Comparison of contact area ratio ($\eta$) obtained from FE models and the corresponding results obtained from the empirical equation: a) at 0.5% rotation, and b) at 1.5% rotation.

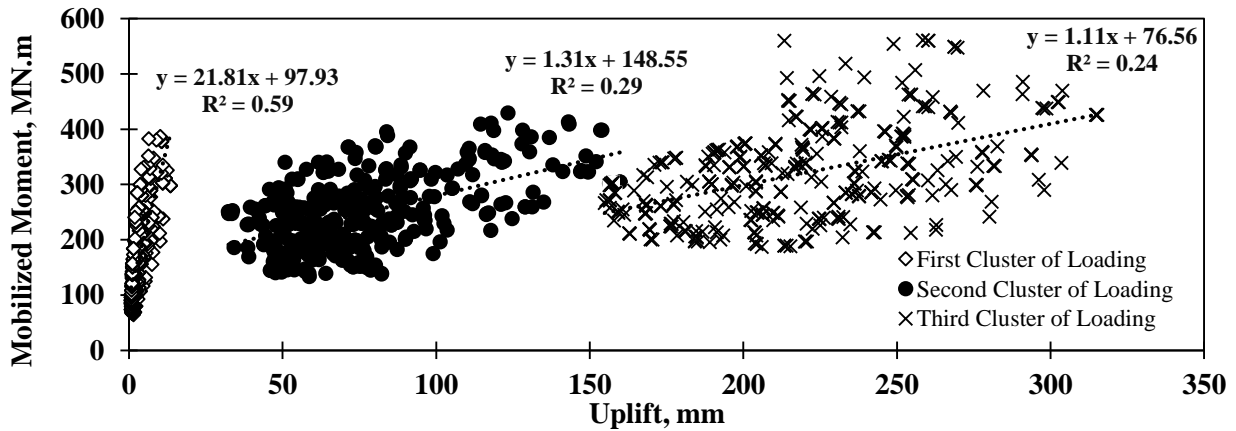

**Figure 18.** Comparison of contact area ratio ($\eta$) obtained from FE models and the corresponding results obtained from the empirical equation: a) at 0.5% rotation, and b) at 1.5% rotation.

As discussed earlier, two main parameters including soil stiffness and structure size (weight and dimension) can affect the dissipated energy of a soil-structure system during rocking. The damping ratio can also be estimated from the rotation-moment hysteresis loops. The following equation accounts for measuring of the damping ratio from the rotation-moment hysteresis loop (see Figure 19):

$$\xi = \frac{1}{4\pi}\left(\frac{Area\ of\ Hysteresis\ loop}{Area\ of\ Traingle\ OAB}\right) \tag{9}$$

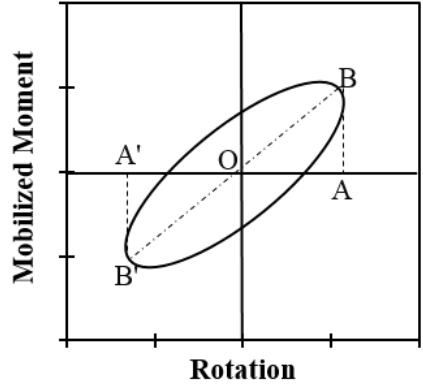

**Figure 19.** Measuring damping ratio using a hysteresis loop

Using the method described in Eq. 9, the maximum damping ratio ($\xi$) was determined for the soil-structure systems at three different clusters of rotations (0.15%, 0.5%, and 1.5%) as demonstrated in Figure 20. It is shown that higher amount of energy is dissipated when a larger amount of rotation is applied to the superstructure.

Figure 20 illustrates the permanent displacement-damping ratio for different amount of foundation rotations. The energy dissipates by yielding of the soil beneath the foundation which is associated with foundation settlement. The damping ratio ($\xi$) increases with an increase in permanent displacement. A summary of descriptive statistics of maximum damping ratio for the studied soil-structure systems at different rotations is illustrated in Figure 21. The average damping ratio was found to be as 6.93, 10.23, and 19.42 for the first, second, and third clusters of loading, respectively. The results show that more variation in damping ratio is observed as the applied rotation increases.

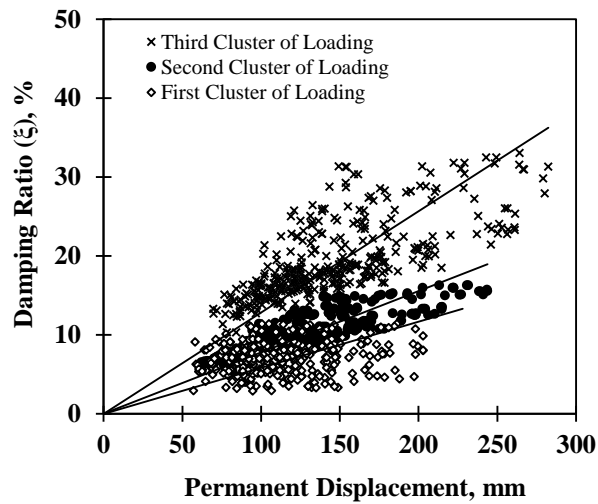

**Figure 20.** Variation of damping ratio at different foundation rotation with the permanent displacement

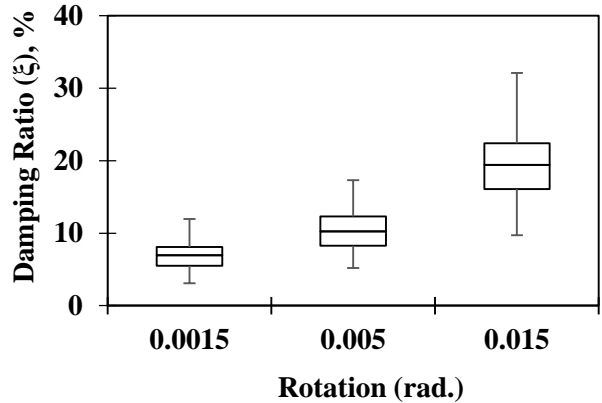

**Figure 21.** Descriptive statistics of damping ratio (ξ) at different rotations.

## 5. Summary and Conclusions

In this paper, the effect of nonlinear properties of geomaterial on the soil-structure interaction during rocking of the structure due to cyclic loading is investigated. A 2D dynamic FE model was assembled to simulate a superstructure imparting energy to the underneath soil at a given amplitude and frequency. To get a better insight into the rocking behavior of the soil-structure system, nonlinear elastic-perfectly plastic behavior was considered for the simulated geomaterial. A set of nonlinear $K_{2(max)}$ parameter was randomly selected for the soil of foundation, from a range of values considered for sandy materials. The foundation-superstructure system was subjected to a slow lateral cyclic loading in three different clusters having three amplitude values. The foundation displacement at the beginning of cyclic loading was recorded and was then compared to the settlements at the end of each cluster of loading. The average permanent displacements for first, second, and third clusters of loading were found to be about 1.11, 1.25, and 1.43 times that of the average initial displacements.

High amplitudes of lateral loading cause a partial separation of the foundation from the supporting geomaterials, and thus, the foundation loses a portion of its initial contact with the soil system. The permanent displacement of the foundation can be related to the foundation separation and the corresponding foundation-soil contact area. The permanent settlement increases with a decrease in the magnitude of uplift. Greater uplift can be observed when geomaterial is stiffer. The structure is more susceptible to lose its contact with the soil of the foundation when it is located on the denser material as compared to a loose one. Therefore, the ratio of $A_c/A$ decreases and foundation loses more contact area as the stiffness of geomaterials increases.

An extensive parametric analysis was carried out to evaluate the amount of the dissipated energy during foundation rocking. The dissipation of energy, indeed, was determined using the rotation-moment hysteresis loops. As judged by the enclosed area of the hysteresis loops, the taller buildings located on softer soils are inclined to dissipate more input-energy in comparison to the smaller buildings sitting on the stiffer material. The rotation-moment hysteresis loops showed that the rotational stiffness degrades as the foundation-superstructure system is subjected to a higher amplitude of lateral loading. The mobilized moment was closely related to the amount of uplift as the mobilized moment increased with the increase in foundation uplift. It was also shown that the higher amount of input-energy is dissipated when the foundation is more in contact with the soil system during rocking.

With all these taken into account, the numerical analyses evidently show that there might be a possibility to use plastic deformation of the soil as the beneficial outcome of a controlled rocking for performance enhancement of soil-structure systems during a strong motion.